\documentclass[9pt,conference]{IEEEtran}
\usepackage{dcase2026}

\usepackage{bm} 

\usepackage{dcase2026,amsmath,graphicx,url,times,booktabs, tabularx}

\usepackage{subcaption}

\title{Reasoning-Oriented Post-Training and Inference-Time LoRA Rescaling for Audio-Dependent Question Answering}

\name{Weiteng Hu$^{1,2}$,
      Yin Cao$^{1}$,
      Jun Yang$^{1,2,3}$
      }
\address{$^1$ Institute of Acoustics, Chinese Academy of Sciences, Beijing, China\\ \{huweiteng, jyang\}@mail.ioa.ac.cn\\
yin.k.cao@gmail.com\\
         $^2$ University of Chinese Academy of Sciences, Beijing, China\\
         $^3$ State Key Laboratory of Acoustics and Marine Information, Institute of Acoustics,\\
        Chinese Academy of Sciences, Beijing, China \\
}

\begin{document}

\maketitle

\begin{abstract}
Audio-Dependent Question Answering (ADQA) requires Large Audio-Language Models (LALMs) to answer questions whose correct answers depend on the given audio content. Successful ADQA requires accurate audio perception, identification of question-relevant evidence, and cross-modal reasoning. Using the official ADQA dataset of DCASE 2026 Task 5, we investigate reasoning-oriented post-training with Low-Rank Adaptation (LoRA) and inference-time LoRA rescaling for both Qwen2.5-Omni and MOSS-Audio-8B-Thinking. We introduce a structured Chain-of-Thought (CoT) framework that decomposes the reasoning process into question analysis, question type, audio evidence, and reasoning. We then analyze how task-specific LoRA adaptation affects the two backbones and further explore inference-time rescaling of trained LoRA adapters. Experiments on the development set reveal markedly backbone-dependent behavior: post-training improves the Qwen-based systems but substantially degrades MOSS-Audio under our supervised fine-tuning configuration. Moderate LoRA rescaling further improves the best Qwen system's top-1 accuracy from 58.93\% to 61.05\% and partially restores the performance of the fine-tuned MOSS-Audio models, while the best MOSS-Audio system achieves 67.70\% top-1 accuracy. Our submitted systems ranked third overall and second among lightweight systems under 10B parameters in the challenge.

\end{abstract}

\begin{IEEEkeywords}
Audio-Dependent Question Answering, Large Audio-Language Model, Chain of Thought Reasoning, Reinforcement Learning, Low-Rank Adaptation
\end{IEEEkeywords}

\section{Introduction}
\label{sec:intro}

Large Audio-Language Models (LALMs) like MOSS-Audio \cite{yang2026moss}, Audio Flamingo \cite{kong2024audio1,ghosh2025audio2,ghosh2026audio3}, and Qwen3-Omni \cite{xu2025qwen3} have recently demonstrated strong capabilities across a range of audio-language tasks. These models can process speech, music, and environmental audio and generate textual outputs for tasks such as automatic speech recognition (ASR), audio captioning, and audio question answering (AQA). Among these tasks, AQA is particularly demanding because the model must understand the textual question, identify the question-relevant information in the audio, and integrate acoustic and linguistic evidence to derive the final answer.

However, strong performance on existing AQA benchmarks does not necessarily demonstrate that a LALM grounds its answers in the provided audio. Questions and candidate answers may contain exploitable linguistic cues, enabling models to rely on textual priors or encoded world knowledge rather than question-relevant acoustic evidence. He et al. \cite{he2026measuring} showed that, on benchmarks such as MMAU \cite{sakshi2025mmau}, MMSU \cite{wang2025mmsu}, and MMAR \cite{ma2026mmar}, LALMs could still answer a subset of questions correctly when the original audio was replaced with silence. This finding suggests that some benchmark samples can be solved without meaningful use of the audio input.

To address this issue, DCASE 2026 Task 5 introduces Audio-Dependent Question Answering (ADQA), a multiple-choice task in which the correct answer is intended to require information from the associated audio. The task uses Audio-Dependency Filtering (ADF) to categorize samples according to the contribution of the audio input, distinguishing weak- and strong-audio-contribution questions. The official training set, AudioMCQ-StrongAC-GeminiCoT, is derived from the strong-audio-contribution subset of AudioMCQ \cite{he2026measuring} and contains Chain-of-Thought (CoT) annotations \cite{wei2022chain} generated by Gemini \cite{team2023gemini}. These strongly audio-dependent samples and reasoning annotations provide supervision for reasoning-oriented post-training in ADQA.

Post-training is widely used to adapt LALMs to specific downstream tasks. Supervised fine-tuning (SFT) can adapt a model to task-specific instructions and output formats, while reinforcement learning (RL) methods such as Group Relative Policy Optimization (GRPO) \cite{shao2024deepseekmath} can further optimize the model using automatically verifiable rewards. In ADQA, answer-level rewards are particularly convenient because the predicted option can be directly compared with the ground-truth answer. However, such outcome-level rewards do not directly assess whether the model has identified question-relevant acoustic evidence or whether its intermediate reasoning is grounded in the audio. Consequently, a model may obtain a high reward while producing ungrounded or degenerate reasoning trajectories.

Task-specific post-training, however, is not uniformly beneficial across models. Although fine-tuning may improve performance on the target task, it can also compromise capabilities acquired during pretraining \cite{luo2025empirical}. SALMONN \cite{tang2024salmonn}, for example, reported that cross-modal instruction tuning could suppress some emergent capabilities and that reducing the inference-time scaling of its Low-Rank Adaptation (LoRA) modules \cite{hu2022lora} could partially recover these capabilities, although with reduced performance on the fine-tuned tasks.
This observation suggests that inference-time adapter strength may control the trade-off between task-specific adaptation and preservation of the base model's original capabilities.

In this work, we investigate reasoning-oriented post-training using Low-Rank Adaptation (LoRA) and inference-time LoRA rescaling for ADQA with Qwen2.5-Omni \cite{xu2025qwen25omnitechnicalreport} and MOSS-Audio-8B-Thinking \cite{yang2026moss}. For Qwen2.5-Omni, we compare two post-training pipelines. The first uses answer-only SFT followed by GRPO, whereas the second applies structured CoT SFT followed by Group Reward-Decoupled Normalization Policy Optimization (GDPO) \cite{liu2026gdpo}. The structured CoT schema decomposes the model output into question analysis, question type, audio evidence, reasoning, and the final answer. For MOSS-Audio-8B-Thinking, we compare its native zero-shot inference with task-specific LoRA-based SFT while preserving its original reasoning format. Finally, for each trained adapter, we vary the LoRA scaling factor during inference to examine how adapter strength affects task-specific adaptation and ADQA performance.

Overall, this paper makes the following contributions:

\begin{itemize}
    \item \textbf{Structured reasoning-oriented post-training.} We develop a structured CoT framework for ADQA that organizes the model output into question analysis, question type, audio evidence, reasoning, and the final answer. This formulation enables explicit supervision of intermediate reasoning fields and field-wise reward design during reinforcement learning.

    \item \textbf{Empirical analysis of backbone-dependent adaptation.} We compare task-specific LoRA post-training on Qwen2.5-Omni and MOSS-Audio-8B-Thinking and observe markedly different adaptation behavior under our experimental settings: post-training improves the Qwen2.5-Omni systems but substantially degrades the fine-tuned MOSS-Audio-8B-Thinking systems.

    \item \textbf{Inference-time LoRA rescaling for ADQA.} We systematically investigate how varying the scaling factor of trained LoRA adapters at inference time affects ADQA performance. The results reveal backbone- and pipeline-dependent behavior: moderate rescaling improves the best Qwen2.5-Omni system, while smaller scaling factors partially recover the SFT-induced performance degradation of MOSS-Audio-8B-Thinking.
    
\end{itemize}

The remainder of this paper is organized as follows. \Cref{sec:method} describes the proposed post-training and inference-time LoRA rescaling methods. \Cref{sec:exp} presents the experimental setup and results. \Cref{sec:conclusion} concludes the paper. Source code is available\footnote{\url{https://github.com/WeitengHu/DCASE2026-Task5}}.

\section{Methodology}
\label{sec:method}
\subsection{Overview}

We conduct post-training and inference-time LoRA rescaling on Qwen2.5-Omni and MOSS-Audio-8B-Thinking. 
Qwen2.5-Omni is evaluated with answer-level and structured-reasoning pipelines, while MOSS-Audio is evaluated with zero-shot inference and further SFT with its native reasoning format.
We additionally rescale the trained LoRA adapters during inference. All baselines, intermediate checkpoints, and rescaled variants are evaluated with pipeline-matched prompts and the same post-processing procedure.

\subsection{Data Preparation}
\label{sec:data}

We use the official DCASE 2026 Task 5 training set, AudioMCQ-StrongAC-GeminiCoT, for all post-training experiments. This dataset is derived from the strong audio-contribution split of AudioMCQ \cite{he2026measuring} and contains 19,480 audio-dependent multiple-choice questions. For the corresponding samples, we extract the structured annotations available in AudioMCQ and convert them into targets for the different post-training pipelines. 
 
For Qwen-CoT, the SFT target contains only the final answer. For Qwen-Structured-CoT, we reorganize the reasoning annotation into four intermediate fields followed by the final answer:

\begin{verbatim}
<question_analysis>...</question_analysis>  
<question_type>...</question_type>  
<audio_evidence>...</audio_evidence>  
<reasoning>...</reasoning>  
\end{verbatim}

These fields represent the analysis of the question, the type of the question, the question-relevant acoustic evidence, and the reasoning based on the audio evidence to derive the final answer. This representation makes the intermediate outputs explicitly supervised and separately rewardable. For the MOSS-Audio SFT, we retain the natural language reasoning annotation rather than imposing the above field structure, and format it using the model's native \texttt{<think>} and \texttt{</think>} tags followed by the final answer.

\subsection{Qwen2.5-Omni Post-Training}
\label{sec:qwen}

We build two post-training pipelines based on Qwen2.5-Omni-7B: Qwen-CoT and Qwen-Structured-CoT. 

\subsubsection{Qwen-CoT}

Qwen-CoT follows a two-stage post-training pipeline. It uses answer-only SFT followed by GRPO to improve ADQA performance.

\textbf{Stage 1: Answer-only Supervised Fine-tuning.} In the first stage, we perform answer-only SFT. We prompt the model to directly generate the answer without any intermediate reasoning. This stage aims to adapt the model to the ADQA multiple-choice instructions and provides a stable initialization for the subsequent GRPO stage.

\textbf{Stage 2: Group Relative Policy Optimization.} In the second stage, we further optimize the model with GRPO \cite{shao2024deepseekmath}. Unlike the answer-only SFT stage, we switch to a CoT-style prompt that instructs the model to think step by step before the final answer. The reward function consists of an accuracy reward and a format reward:
\begin{equation}
R_{\mathrm{CoT}} =
w_{\mathrm{acc}} r_{\mathrm{acc}} +
w_{\mathrm{fmt}} r_{\mathrm{fmt}},
\end{equation}
where $r_{\mathrm{acc}}=1$ if the extracted answer matches the ground-truth answer and $r_{\mathrm{fmt}}=1$ if the response follows the required output format. This design encourages the model to generate step-by-step reasoning before producing the final answer, without explicitly supervising the intermediate reasoning content. 

\subsubsection{Qwen-Structured-CoT}

Qwen-Structured-CoT also follows a two-stage post-training pipeline, but differs from Qwen-CoT in both the supervised target and the reward design.

\textbf{Stage 1: Structured CoT Supervised Fine-tuning.} In the first stage, we perform structured CoT SFT using the reconstructed annotations described in \cref{sec:data}. Instead of directly predicting only the final answer, the model is trained to generate the entire structured reasoning sequence, including question analysis, question type, audio evidence, reasoning, and the final answer. This stage teaches the model to follow the structured output schema and serves as a stable warm-up for the subsequent reinforcement learning.

\textbf{Stage 2: Group Reward-Decoupled Normalization Policy Optimization.} In the second stage, we design a composite reward that evaluates both the final answer and intermediate reasoning fields. For each prompt, let $\{y_i\}_{i=1}^{G}$ denote the $G$ sampled responses. The reward for response $y_i$ is composed of

\begin{equation}
\label{eq:gated_rewards}
\begin{aligned}
\mathbf{r}^{i} = \big(
& r_{\mathrm{acc}}^i,\,
r_{\mathrm{fmt}}^i,\,
r_{\mathrm{qtype}}^i,\,
r_{\mathrm{qtype}}^i s_{\mathrm{qa}}^i, \\
& r_{\mathrm{acc}}^i s_{\mathrm{ae}}^i,\,
r_{\mathrm{acc}}^i s_{\mathrm{reason}}^i,\,
r_{\mathrm{acc}}^i r_{\mathrm{len}}^i
\big).
\end{aligned}
\end{equation}

Here, $r_{\mathrm{acc}}^i=1$ if the extracted answer matches the ground-truth answer, $r_{\mathrm{fmt}}^i=1$ if the model's reasoning follows the structured schema described in \cref{sec:data}, $r_{\mathrm{qtype}}^i=1$ if the predicted question type matches the reference question type, and $r_{\mathrm{len}}^i=1$ if the length of the response falls within a predefined valid range. 

$s_{\mathrm{qa}}^i$, $s_{\mathrm{ae}}^i$, and $s_{\mathrm{reason}}^i$ denote the cosine similarity calculated by Qwen3-Embedding-0.6B \cite{zhang2025qwen3} for question analysis, audio evidence, and reasoning. For each textual field, $f\in\{\mathrm{qa},\mathrm{ae},\mathrm{reason}\}$, the similarity is defined as
\begin{equation}
s_{f}^i =
\cos\!\left(
E({y}_{f}^{i}), E(\hat{y}_{f})
\right),
\label{eq:similarity_reward}
\end{equation}
where $E(\cdot)$ denotes Qwen3-Embedding-0.6B \cite{zhang2025qwen3}, ${y}_{f}^{i}$ is the generated field, and $\hat{y}_{f}$ is its reference annotation. 

To reduce spurious rewards for incorrect reasoning trajectories, we apply a gated reward design as shown in \cref{eq:gated_rewards}. Specifically, the question-analysis similarity is applied only when the prediction of question type is correct ($r_{\mathrm{qtype}}=1$), while the audio-evidence similarity, reasoning similarity, and length regularization reward are gated by $r_{\mathrm{acc}}$. This gated design prevents the model from receiving high rewards for intermediate reasoning fields when the predicted question type or final answer is incorrect. To avoid reward collapse in standard GRPO, we adopt GDPO \cite{liu2026gdpo} in this stage, which normalizes each reward separately before aggregation. For each reward $r_m^i$, GDPO first computes a group-normalized advantage:
\begin{equation}
a_m^i =\frac{r_m^i-\mu_m}{\sigma_m+\epsilon},\quad
\mu_m=\frac{1}{G}\sum_{i=1}^{G}r_m^i,
\label{eq:gdpo_group}
\end{equation}
where $\sigma_m$ is the standard deviation of reward $m$ over the $G$ responses. The overall advantage is then obtained by summing the weighted normalized advantages across all objectives:

\begin{equation}
\widetilde{A}_i=\sum_{m}w_m a_m^i,
\label{eq:gdpo_aggregation}
\end{equation} 

where $w_m$ is the weight of each normalized advantage. These weights are set empirically and their values are reported in \cref{sec:setup}.

\subsection{MOSS-Audio-8B-Thinking Systems}

Unlike Qwen2.5-Omni-7B, MOSS-Audio-8B-Thinking is a native thinking model. We mainly preserve its original reasoning behavior through zero-shot inference, and further perform LoRA-based SFT.

\subsubsection{Zero-shot inference}
We use the native thinking format of MOSS-Audio-8B-Thinking and only control the final-answer format. Specifically, MOSS-Thinking-Full asks the model to produce the complete answer text after reasoning, while MOSS-Thinking-Label constrains the final answer to be only the option label. The label-only setting is designed to reduce answer-mapping ambiguity during post-processing.

\subsubsection{LoRA-based SFT}
We further perform LoRA-based SFT on MOSS-Audio-8B-Thinking. Since MOSS-Audio-8B-Thinking already has a native thinking format, we do not impose the external structured schema used in Qwen-Structured-CoT. Instead, we preserve the natural language reasoning format and organize it according to the native thinking format of MOSS-Audio-8B-Thinking with \texttt{<think>} and \texttt{</think>} tags. This target format reduces the mismatch between supervised responses and the model's native output schema.

\subsection{Inference-Time LoRA Rescaling}
\label{sec:lora_rescaling}
Given a frozen pretrained weight $W_0 \in \mathbb{R}^{d \times k}$, LoRA introduces a low-rank update $\Delta W = BA$, where $B \in \mathbb{R}^{d \times r}$ and $A \in \mathbb{R}^{r \times k}$ are trainable low-rank matrices. The adapted weight is computed as:
\begin{equation}
W = W_0 + \frac{\alpha}{r} BA = W_0 + \gamma BA,
\end{equation}
where $r$ is the LoRA rank, $\alpha$ is the LoRA alpha, and $\gamma = \frac{\alpha}{r}$ is the scaling factor of the LoRA adapter. 

Inference-time LoRA rescaling refers to varying $\alpha$, equivalently $\gamma$, during inference while keeping the trained LoRA matrices fixed. This operation changes only the scaling factor of the already trained LoRA adapters and involves no additional training.

\subsection{Post-Processing}
\label{sec:post}
All evaluated systems share the same post-processing pipeline. First, we extract the final answer from the model response and then map it to an option using exact or containment string matching. If no answer is extracted or no valid option is matched, we employ Qwen3-Embedding-0.6B \cite{zhang2025qwen3} to select the candidate choice that is the most semantically similar to the model response. Furthermore, to mitigate option-order bias and enhance robustness, inference is performed across both the original choice order and four randomly shuffled choice orders for each system. The final answer is determined by majority voting over these five runs.

\section{Experiments}
\label{sec:exp}

\subsection{Experimental Setup}
\label{sec:setup}
All experiments were conducted on 4 NVIDIA GeForce RTX 4090 GPUs using bfloat16 mixed-precision training and all LoRA-trained variants used rank $r=8$ and
$\alpha_{\mathrm{train}}=32$, corresponding to $\gamma=4$. We applied LoRA to all linear layers in the audio encoder, multimodal projector, and language model. 

For the Qwen2.5-Omni systems, all stages were implemented within the ms-swift \cite{zhao2025swift} framework. The SFT stage was trained for one epoch with a learning rate of $1\times10^{-4}$, per-device batch size of 1, and gradient accumulation steps of 8. The RL stage was trained for one epoch with a learning rate of $1\times10^{-5}$, a KL regularization coefficient of $\beta=0.001$, number of generations of 8, a generation batch size of 32, and a maximum completion length of 512. The reward weights were selected empirically. We used $w_{\mathrm{acc}}=2.0$, $w_{\mathrm{fmt}}=0.5$ for Qwen-CoT, and
$w_{\mathrm{acc}}=2.0$, $w_{\mathrm{fmt}}=0.5$, $w_{\mathrm{qtype}}=0.5$, $w_{\mathrm{qa}}=0.25$, $w_{\mathrm{ae}}=0.5$, $w_{\mathrm{reason}}=0.5$, $w_{\mathrm{len}}=0.25$ for Qwen-Structured-CoT. Greedy decoding was used for all Qwen2.5-Omni systems.

For the MOSS-Audio-8B-Thinking systems, the SFT stage was implemented with Hugging Face Transformers \cite{wolf2020transformers} and Accelerate. The model was trained for one epoch with a learning rate of $1\times10^{-4}$, per-device batch size of 1, and gradient accumulation steps of 8. During inference, we used a temperature of 1.0, top-$p$ of 1.0, and top-$k$ of 50 for all MOSS-Audio-8B-Thinking systems.

\begin{figure*}[!t]
    \centering
    \begin{subfigure}[t]{0.24\textwidth}
        \centering
        \includegraphics[width=\linewidth]{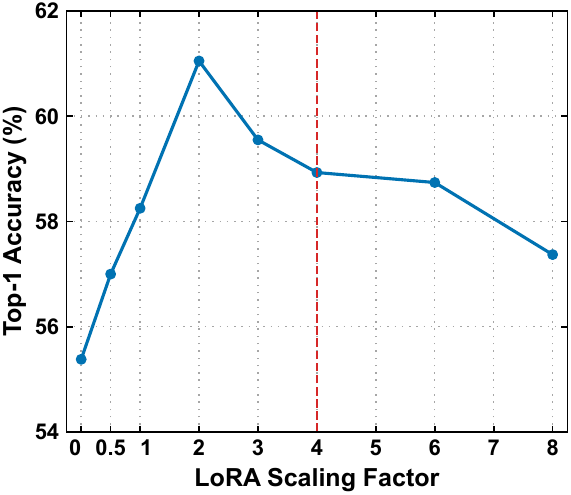}
        \caption{Qwen-CoT}
        \label{fig:lora_qwen_cot}
    \end{subfigure}
    \hfill
    \begin{subfigure}[t]{0.24\textwidth}
        \centering
        \includegraphics[width=\linewidth]{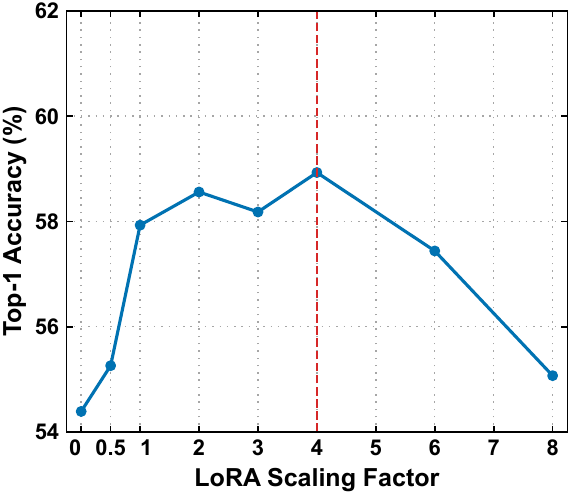}
        \caption{Qwen-Structured-CoT}
        \label{fig:lora_qwen_struct}
    \end{subfigure}
    \hfill
    \begin{subfigure}[t]{0.24\textwidth}
        \centering
        \includegraphics[width=\linewidth]{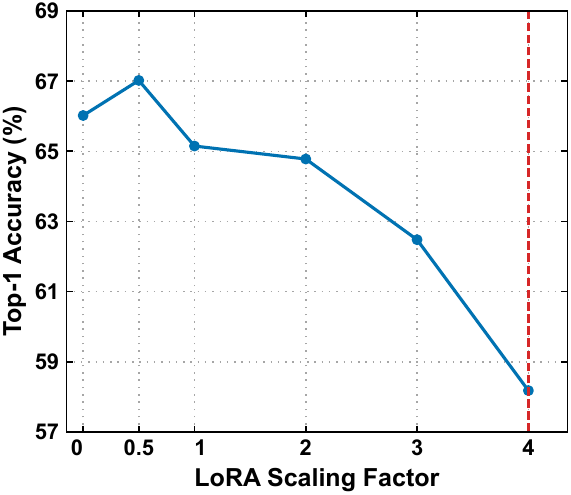}
        \caption{MOSS-Thinking-Full}
        \label{fig:lora_moss_full}
    \end{subfigure}
    \hfill
    \begin{subfigure}[t]{0.24\textwidth}
        \centering
        \includegraphics[width=\linewidth]{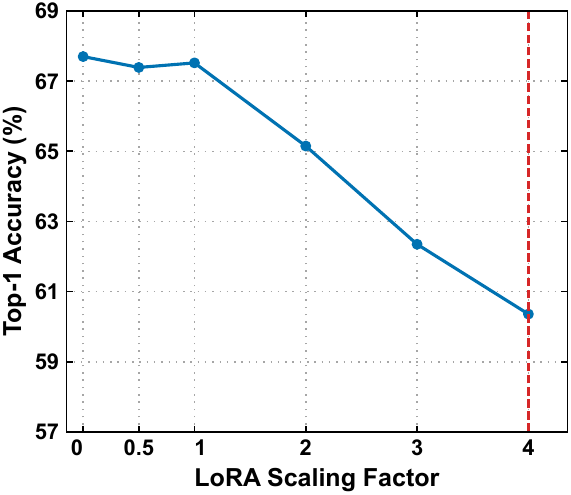}
        \caption{MOSS-Thinking-Label}
        \label{fig:lora_moss_label}
    \end{subfigure}
    \caption{Performance changes of different systems on the development set when varying the LoRA scaling factor during inference. The red dashed line indicates the LoRA scaling factor during training.}
    \label{fig:lora}
\end{figure*}

\subsection{Effect of Structured Reasoning Post-Training}
\label{sec:reasoning_results}

All models are evaluated on the DCASE 2026 Task 5 development set. \Cref{tab:main_results} reports the top-1 accuracy of our systems and the intermediate training stages. Each system is evaluated using its own pipeline-specific inference prompt, which is kept fixed across the corresponding baseline, SFT, and RL checkpoints. 

\begin{table}[ht]
\centering
\caption{Top-1 accuracy (\%) of different systems and training stages on the development set. $\Delta$ compares the final trained checkpoint with the prompt-matched base model.}
\label{tab:main_results}
\scriptsize
\renewcommand{\arraystretch}{1.2}
\begin{tabular}{lcccc}
\toprule
\textbf{System} & \textbf{Baseline} & \textbf{SFT} & \textbf{RL} & $\boldsymbol{\Delta}$ \\
\midrule
Qwen-CoT              & 55.38 & 56.50 & 58.93 & $+$3.55 $\uparrow$ \\
Qwen-Structured-CoT   & 54.39 & 57.93 & 58.93 & $+$4.54 $\uparrow$ \\
\midrule
MOSS-Thinking-Full    & 66.02 & 58.18 & --    & $-$7.84 $\downarrow$ \\
MOSS-Thinking-Label   & 67.70 & 60.36 & --    & $-$7.34 $\downarrow$ \\
\bottomrule
\end{tabular}
\end{table}

For the Qwen2.5-Omni systems, both post-training strategies improve over the baseline performance. For Qwen-CoT, the answer-only SFT stage slightly improves the baseline from 55.38\% to 56.50\%, and the subsequent GRPO stage further increases the accuracy to 58.93\%. Qwen-Structured-CoT exhibits a larger gain after the structured CoT SFT stage, improving the accuracy from 54.39\% to 57.93\%. After GDPO optimization with the structured reward function, the final accuracy reaches 58.93\%, which is the same as the final Qwen-CoT system.

These results show that structured reasoning supervision produces a larger accuracy gain during the SFT stage, while the subsequent RL stages narrow the performance gap between the two pipelines. One possible explanation is that explicit supervision over question analysis, question type, audio evidence, and reasoning is particularly useful for adapting a backbone without a native thinking format. This contrast suggests that the two pipelines distribute their adaptation gains differently across training stages: structured supervision contributes most of its benefit during SFT, while Qwen-CoT relies more heavily on RL to develop a reasoning-oriented response pattern.



\subsection{Backbone-Dependent Post-Training Behavior}
\label{sec:backbone_results}

As shown in \cref{tab:main_results}, MOSS-Audio-8B-Thinking exhibits the opposite trend to Qwen2.5-Omni. Its zero-shot systems achieve 66.02\% and 67.70\% top-1 accuracy, whereas LoRA SFT reduces them to 58.18\% and 60.36\%, corresponding to losses of 7.84 and 7.34 points. 

The opposite adaptation trends of the two backbones suggest that task-specific SFT is not uniformly beneficial for different models. Qwen2.5-Omni starts from a lower baseline and benefits from explicit task adaptation, whereas MOSS-Audio-8B-Thinking achieves superior zero-shot performance but degrades sharply after SFT.
One possible explanation is that the task-specific adapter disturbs pretrained behaviors of MOSS-Audio-8B-Thinking under our training configuration. This observation motivates the adapter rescaling in \cref{sec:eff_rescale}, where we further vary the LoRA scaling factor during inference for both Qwen2.5-Omni and MOSS-Audio-8B-Thinking.




\subsection{Inference-Time LoRA Rescaling Analysis}
\label{sec:eff_rescale}
This section explores the influence of the LoRA scaling factor during inference without additional fine-tuning. Following the procedure in \Cref{sec:lora_rescaling}, we keep all model and adapter parameters fixed and sweep the inference-time scaling factor $\gamma$. \Cref{fig:lora} presents the complete scaling curves, and \Cref{tab:lora_best} reports the best observed nonzero scaling factor for each system. Here, $\gamma=0$ disables the adapter, whereas $\gamma=4$ is the default scaling used during training.

\begin{table}[h]
\centering
\caption{Best observed top-1 accuracy (\%) obtained among the tested nonzero inference-time LoRA scaling factors on the development set. Base and Default denote top-1 accuracy of $\gamma=0$ and the training-time value $\gamma=4$, respectively.}
\label{tab:lora_best}
\scriptsize
\setlength{\tabcolsep}{2.6pt}
\renewcommand{\arraystretch}{1.15}
\begin{tabular}{lcccc}
\toprule
\textbf{System} & \textbf{Baseline} & \textbf{Default} & \textbf{Best Observed $\boldsymbol{\gamma}$} & \textbf{Best Acc.}\\
\midrule
Qwen-CoT        & 55.38 & 58.93 & 2.0 & \textbf{61.05} \\
Qwen-Structured-CoT & 54.39 & \textbf{58.93} & 4.0 & \textbf{58.93} \\
MOSS-Thinking-Full       & 66.02 & 58.18 & 0.5 & \textbf{67.02} \\
MOSS-Thinking-Label      & \textbf{67.70} & 60.36 & 1.0 & 67.52 \\
\bottomrule
\end{tabular}
\end{table}

As shown in \cref{fig:lora}, both Qwen-based systems exhibit non-monotonic trends as the LoRA scaling factor increases, with accuracy improving at moderate scaling values but degrading under overly large adapter strength. For Qwen-CoT, moderately reducing the scaling factor from the training value improves performance, and the best performance among the tested values is obtained at $\gamma=2$, where the top-1 accuracy reaches 61.05\%. This suggests that the default LoRA scaling during training is not always optimal at inference, and a smaller adapter contribution can yield better task-specific performance.
For Qwen-Structured-CoT, the training scaling $\gamma=4$ remains the best tested configuration, achieving 58.93\% top-1 accuracy. One possible explanation is that the structured CoT system relies more strongly on the learned adapter to maintain its structured response schema. However, overly large scaling can still hurt performance.

For MOSS-Audio-8B-Thinking systems, reducing the LoRA scaling primarily serves as a recovery mechanism against fine-tuning degradation. The training LoRA SFT setting with $\gamma=4$ leads to a substantial degradation compared with zero-shot inference. For MOSS-Thinking-Full, a small scaling factor of $\gamma=0.5$ leads to a slight improvement, increasing the top-1 accuracy from the zero-shot result of 66.02\% to 67.02\%. For MOSS-Thinking-Label, the best observed nonzero configuration reaches 67.52\% at $\gamma=1$, but remains below the adapter-disabled result of 67.70\%. These results show that weakening the trained adapter can recover much of the ADQA performance lost under the default SFT scaling, although it does not consistently outperform the corresponding base model.

Overall, the sweep of scaling factor reveals that inference-time LoRA strength has markedly different effects across systems. Moderate rescaling improves Qwen-CoT, while the training-time value remains the best tested setting for Qwen-Structured-CoT. For MOSS-Audio, reduced scaling largely recovers the SFT-induced performance drop. Since this rescaling is performed only at inference without additional fine-tuning, it provides a simple and low-cost way to recover, and in some cases further improve, ADQA performance in the evaluated LoRA-adapted systems.

\section{Conclusion}
\label{sec:conclusion}
In this paper, we investigated reasoning-oriented LoRA post-training and inference-time LoRA rescaling for Audio-Dependent Question Answering using Qwen2.5-Omni and MOSS-Audio-8B-Thinking. For Qwen2.5-Omni, both post-training pipelines improve over the baseline. Structured CoT yields a larger improvement during SFT with explicit supervision over question analysis, question type, audio evidence, and reasoning, whereas the subsequent RL stages narrow the final performance gap between the two pipelines. In contrast, MOSS-Audio-8B-Thinking exhibits strong zero-shot performance, but task-specific SFT substantially degrades its accuracy. 
We further analyze inference-time LoRA rescaling and find that adjusting the scaling factor of a trained LoRA adapter at inference time without additional training can recover much of the performance loss introduced by SFT. Overall, post-training is not uniformly beneficial across LALMs. The results suggest that reasoning supervision, backbone choice, and adapter strength should be jointly considered when adapting models to ADQA.

\clearpage
\bibliographystyle{IEEEtran}
\bibliography{refs}







\end{document}